\documentclass[reprint,
superscriptaddress,
 amsmath,amssymb,
 aps,
 apl,
]{revtex4-1}

\usepackage{graphicx}
\usepackage{dcolumn}
\usepackage{bm}
\usepackage{xcolor} 
\usepackage{filecontents}

\begin{document}


\title{Isotopic enrichment of silicon by high fluence $^{28}$Si$^-$ ion implantation}

\author{D. Holmes}
\email[Corresponding author: ]{dholmes1@student.unimelb.edu.au}
\affiliation{Centre for Quantum Computing and Communication Technology, School of Physics, The University of Melbourne,
Melbourne, VIC 3010, Australia}
\author{B. C. Johnson}
\affiliation{Centre for Quantum Computing and Communication Technology, School of Physics, The University of Melbourne,
Melbourne, VIC 3010, Australia}
\author{C. Chua}
\affiliation{Silicon Quantum Computing, Sydney, NSW 2052, Australia}
\affiliation{School of Physics, University of New South Wales Sydney, Sydney, NSW 2052, Australia}
\author{B. Voisin}
\affiliation{Silicon Quantum Computing, Sydney, NSW 2052, Australia}
\affiliation{School of Physics, University of New South Wales Sydney, Sydney, NSW 2052, Australia}
\author{S. Kocsis}
\affiliation{Silicon Quantum Computing, Sydney, NSW 2052, Australia}
\affiliation{School of Physics, University of New South Wales Sydney, Sydney, NSW 2052, Australia}
\author{S. Rubanov}
\affiliation{Advanced Microscopy Facility, Bio21 Institute, The University of Melbourne, VIC 3010, Australia}
\author{S. G. Robson}
\affiliation{Centre for Quantum Computing and Communication Technology, School of Physics, The University of Melbourne,
Melbourne, VIC 3010, Australia}
\author{J. C. McCallum}
\affiliation{Centre for Quantum Computing and Communication Technology, School of Physics, The University of Melbourne,
Melbourne, VIC 3010, Australia}
\author{D. R. McCamey}
\affiliation{ARC Centre of Excellence in Exciton Science, School of Physics, University of New South Wales Sydney, NSW 2052, Australia}
\author{S. Rogge}
\affiliation{Centre for Quantum Computation and Communication Technology, School of Physics, University of New South Wales Sydney, Sydney, NSW 2052, Australia}
\author{D. N. Jamieson}
\affiliation{Centre for Quantum Computing and Communication Technology, School of Physics, The University of Melbourne,
Melbourne, VIC 3010, Australia}


\begin{abstract}
Spins in the `semiconductor vacuum' of silicon-28 ($^{28}$Si) are suitable qubit candidates due to their long coherence times. An isotopically purified substrate of $^{28}$Si is required to limit the decoherence pathway caused by magnetic perturbations from surrounding $^{29}$Si nuclear spins ($I=1/2$), present in natural Si ($^\text{nat}$Si) at an abundance of 4.67\%. We isotopically enrich surface layers of $^\text{nat}$Si by sputtering using high fluence $^{28}$Si$^-$ implantation. Phosphorus (P) donors implanted into one such $^{28}$Si layer with $\sim$3000 ppm $^{29}$Si, produced by implanting 30 keV $^{28}$Si$^-$ ions at a fluence of $4\times10^{18}$ cm$^{-2}$, were measured with pulsed electron spin resonance, confirming successful donor activation upon annealing. The mono-exponential decay of the Hahn echo signal indicates a depletion of $^{29}$Si. A coherence time of $T_2=285\pm14$ $\mu$s is extracted, which is longer than that obtained in $^\text{nat}$Si for similar doping concentrations and can be increased by reducing the P concentration in future. The isotopic enrichment was improved by employing one-for-one ion sputtering using 45 keV $^{28}$Si$^-$ implantation. A fluence of $2.63\times10^{18}$ cm$^{-2}$ $^{28}$Si$^-$ ions were implanted at this energy into $^\text{nat}$Si, resulting in an isotopically enriched surface layer $\sim$100 nm thick; suitable for providing a sufficient volume of $^{28}$Si for donor qubits implanted into the near-surface region. We observe a depletion of $^{29}$Si to 250 ppm as measured by secondary ion mass spectrometry. The impurity content and the crystallization kinetics via solid phase epitaxy are discussed. The $^{28}$Si layer is confirmed to be a single crystal using transmission electron microscopy. This method of Si isotopic enrichment shows promise for incorporating into the fabrication process flow of Si spin qubit devices. 
\end{abstract}
                           
\maketitle

\section{Spins in $^{28}$Si}

Donor and quantum dot spin qubits in silicon (Si) are attractive candidates for high-fidelity scalable quantum computing architectures \cite{kane1998silicon,loss1998quantum,zwanenburg2013silicon}. Si provides a desirable matrix for hosting spin qubits due to its important role in the microelectronics industry, weak spin-orbit coupling and the existence of isotopes with zero nuclear spin. Natural Si consists of three isotopes: $^{28}$Si (92.23\%), $^{29}$Si (4.67\%) and $^{30}$Si (3.1\%) \cite{de2004handbook}. The dominant source of qubit decoherence in $^\text{nat}$Si is due to coupling with the surrounding $^{29}$Si nuclei, which possess a nuclear spin of $I=1/2$. Dipolar fluctuations of $^{29}$Si spins cause perturbations in the local magnetic field, resulting in a time-varying qubit resonance frequency \cite{de2003theory,witzel2010electron}. This spectral diffusion limits the spin coherence time to around 200 $\mu$s for electrons \cite{pla2012single} and 60 ms for ionised donor nuclei \cite{pla2013high}, as measured for a single phosphorus (P) donor at low temperature using the Hahn-echo pulse sequence. 

\begin{figure}[h]
\centering
\includegraphics[scale=0.26]{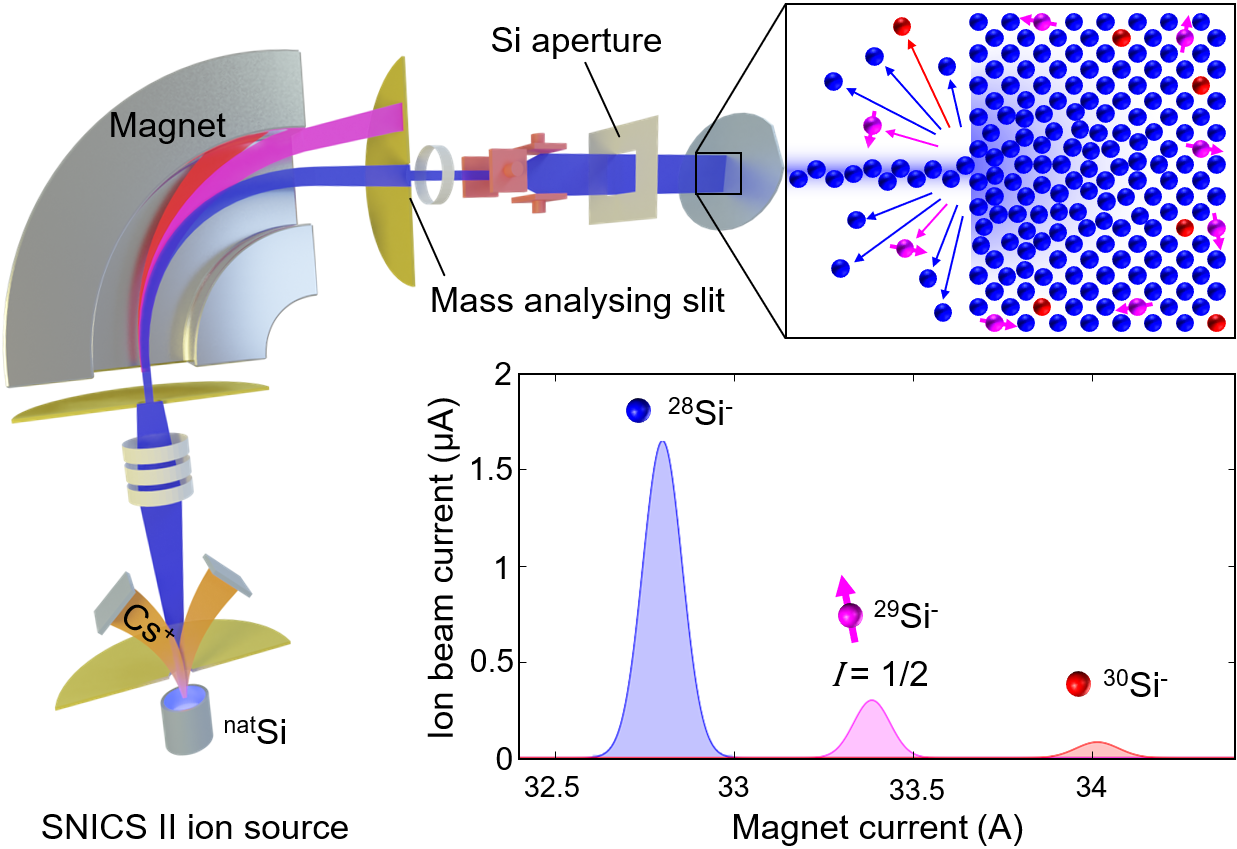}
\caption{A $^{28}$Si$^-$ ion beam, filtered by a mass-selecting magnet from a solid $^\text{nat}$Si source, is used to isotopically enrich a $^\text{nat}$Si substrate surface layer by sputtering. The mass spectrum of the ion implanter shows isotopic resolution of Si.}
\label{fig:implanter}
\end{figure}

Fortunately, $^{28}$Si has no nuclear spin and can therefore provide an ideal low-noise environment for spin qubits. Long coherence times for donor spin qubits in a $^{28}$Si epilayer with 800 ppm residual $^{29}$Si \cite{itoh2014isotope} have been demonstrated, with Hahn-echo decay times of around 1 ms for electrons and 1.75 s for ionised donor nuclei measured for a single P donor at low temperature, which can be further extended with dynamical decoupling \cite{muhonen2014storing}. Isotope engineering of semiconductor materials also has applications for increased thermal conductivity \cite{capinski1997thermal,ruf2000thermal, inyushkin2018ultrahigh}, capable of improved heat dissipation in Si integrated circuits \cite{kizilyalli2005power}.

Isotopically enriched $^{28}$Si can be produced by various methods, many of which involve the centrifugation of silicon tetrafluoride gas to produce high purity $^{28}$SiF$_4$ \cite{takyu1999growth,bulanov2000highly,itoh2003high,becker2009avogadro,becker2010enrichment,sennikov2012towards,abrosimov2017new}. In the Avogadro Project \cite{becker2010enrichment}, $^{28}$SiF$_4$ is converted into isotopically pure silane gas, $^{28}$SiH$_4$, which is used to grow polycrystalline $^{28}$Si by chemical vapour deposition (CVD). Float-zone growth is then used to produce $^{28}$Si single crystal rods. An isotopic purity of $<10$ ppm $^{29}$Si and the highest chemical purity to date ($<4\times10^{15}$ cm$^{-3}$ for C and $<4\times10^{14}$ cm$^{-3}$ for O) was achieved \cite{abrosimov2017new}. 
Epilayers of $^{28}$Si can be grown on $^\text{nat}$Si substrates by CVD \cite{mazzocchi201999,sabbagh2019quantum} (a method employed by the Isonics Corporation \cite{itoh2014isotope}) or molecular beam epitaxy \cite{kojima2003growth,bracht2012reduced,nakabayashi2000epitaxial} (suitable for encapsulation of scanning tunneling microscopy-placed donors in $^{28}$Si \cite{o2001towards,oberbeck2002encapsulation}). These methods rely on isotopically enriched sources of silane gas or solid-state Si. Mass spectrometry, on the other hand, can be used to separate $^{28}$Si ions from a natural silane gas source. This has been used in conjunction with hyperthermal energy ion beam deposition to achieve a residual $^{29}$Si fraction of $<1$ ppm in a layer of $^{28}$Si \cite{dwyer2014enriching,tang2020targeted}. However, the concentration of C and O was greater than $1\times10^{19}$ cm$^{-3}$ \cite{dwyer2014enriching}.
The epitaxial growth of $^{28}$Si thin films by ion beam deposition with a solid $^\text{nat}$Si source has also been achieved by depositing low energy ($\sim$40 eV) $^{28}$Si$^-$ ions \cite{tsubouchi2001epitaxial}.

In this work, we achieve isotopic enrichment by implanting a high fluence of $^{28}$Si$^-$ ions, mass-separated from an accelerated ion beam produced from a solid-state $^\text{nat}$Si source, into a $^\text{nat}$Si substrate, shown schematically in Fig. \ref{fig:implanter}. Since any isotope in the substrate can be sputtered from the surface, but only $^{28}$Si ions are implanted, the levels of $^{29}$Si and $^{30}$Si are depleted with increasing fluence. An enriched surface layer of $^{28}$Si can be produced with sufficient volume for donor qubits; typically implanted to a depth of around 20 nm below the Si surface for effective control and readout by surface nanocircuitry \cite{tosi2017silicon}. This method has the advantage of using standard ion beam laboratory equipment, enabling the integration of \textit{in-situ} enrichment with existing ion-implanted donor qubit fabrication \cite{van2015single}. Additionally, the creation of an amorphous $^{28}$Si layer increases the placement precision of implanted donors by suppressing ion channeling \cite{schreutelkamp1991pre} and can increase the donor electrical activation yield \cite{holmes2019activation}. Solid phase epitaxy (SPE) of an amorphous Si ($a$-Si) surface layer formed by Si implantation can produce near-perfect single-crystal Si ($c$-Si) \cite{zhu1999structural} with a smooth surface \cite{zhu1998surface}, in which no long-range atomic displacement occurs \cite{radek2015atomic}. A one-for-one replacement regime resulting in a planar Si surface (suitable for post-fabrication of nanocircuitry) could allow for economical local enrichment of regions using a focused $^{28}$Si ion beam into which donors are implanted. Minimising the number of $^{29}$Si nuclei within the Bohr radius of the donor electron ($\sim$2 nm for $^{31}$P in Si \cite{kohn1955theory}) reduces the coupling of the donor electron to the dynamics of the $^{29}$Si spin-bath through the contact hyperfine interaction \cite{witzel2010electron}. 

The concentration of impurities introduced to the enriched layer during high fluence $^{28}$Si implantation must be minimised, with particular care taken to avoid co-implantation of the molecular isobars CO and N$_2$. The presence of C, N and O in the $a$-Si layer at levels of around 0.5 at. \% cause retardation of the SPE regrowth rate \cite{kennedy1977influence} and can lead to polycrystalline nucleation during annealing at increased implant concentrations \cite{williams1983solid}. 
The use of a negative $^{28}$Si ion beam ensures a negligible component of the isobars CO and N$_2$, since these negatively charged molecular ions are electronically unstable.

In the present work, 30-45 keV $^{28}$Si$^-$ ions were implanted at a high fluence into $^\text{nat}$Si. The crystallisation kinetics were determined using time resolved reflectivity (TRR) and the coherence time of P donors implanted into the enriched $^{28}$Si layer was measured using pulsed electron spin resonance (ESR). The extent of $^{29}$Si depletion and impurity levels introduced were measured using secondary ion mass spectrometry (SIMS). The concentration of impurities was sufficiently low to allow for crystallisation to take place via SPE, resulting in a single crystal layer of $^{28}$Si, as shown by transmission electron microscopy (TEM). 

\section{Experiment}

A 150 keV ion implanter, equipped with a SNICS II ion source and a 90$^\circ$ double focusing magnet, was used for all implants. Fig. \ref{fig:implanter} schematically shows the mass spectrum of this implanter with a $^\text{nat}$Si source, demonstrating the mass resolution of the Si isotopes. Near the start of an implant run, the ion beam currents for $^{29}$Si$^-$ and $^{30}$Si$^-$ are higher than their natural abundance as they contain a significant fraction of $^{28}$SiH$^-$ and $^{29}$SiH$^-$, respectively. $^{28}$Si$^-$ implants were performed at room temperature and with a 7$^{\circ}$ tilt off the incident beam axis to suppress ion channeling. A Si aperture, prepared from a wafer of $^\text{nat}$Si, was used to collimate the beam and prevent contamination from forward recoils of foreign atoms. Substrates were given a degreasing clean and a HF etch to remove surface hydrocarbons and native SiO$_2$ before entering the implant chamber to further reduce contamination. A vacuum of less than $1 \times 10^{-7}$ Torr was maintained in the target chamber with a cryopump to reduce impurity incorporation from residual gas by ion bombardment.

A highly intrinsic (4-10 k$\Omega$.cm) float-zone uniform high purity $^\text{nat}$Si (UHPS Topsil) substrate was implanted with 30 keV $^{28}$Si$^-$ at a fluence of $4\times10^{18}$ cm$^{-2}$ followed by $^{31}$P$^-$ implantation (30 keV, $6.5\times10^{11}$ cm$^{-2}$ then 10 keV, $1.5\times10^{11}$ cm$^{-2}$), resulting in a P concentration of $<1.4\times10^{17}$ cm$^{-3}$ throughout the enriched $^{28}$Si layer. To investigate the crystallisation kinetics of the $a$-Si layer, TRR \cite{olson1988kinetics} was used with a laser wavelength of $\lambda=632.8$ nm during an anneal in air at 609 $^{\circ}$C. The rate of crystallisation was compared to an $a$-Si standard: n-type $^\text{nat}$Si amorphised with a much lower fluence of $^{28}$Si$^-$ ions with the following implantation scheme: (0.5 MeV, $3\times10^{15}$ cm$^{-2}$), then (1 MeV, $1\times10^{15}$ cm$^{-2}$) and finally (2 MeV, $1\times10^{15}$ cm$^{-2}$). The SPE growth rate of this $a$-Si standard during the initial stages of the anneal was used to calibrate the temperature of the TRR hot plate. Refractive indices of $n_c=4.086$ for $c$-Si \cite{jellison1994optical} and $n_a=4.831$ for $a$-Si \cite{olson1988kinetics} were used for the SPE rate calculation.

Pulsed ESR was performed on P donors implanted in the $^{28}$Si layer of the $^{28}$Si$^-$ (30 keV, $4\times10^{18}$ cm$^{-2}$) implanted sample after SPE. The sample was mounted onto the surface of a superconducting cavity made by dry etching a 100 nm-thick NbTiN film \cite{kobayashi2020engineering,weichselbaumer2019quantitative}. The sample, along with the cavity, was mounted on the mixing chamber of a dilution refrigerator, with a base temperature of $\sim$16 mK. Pulses were sent to the cavity using a vector source at the resonant frequency of the cavity ($f=6.028$ GHz), and the detected echo signal was then pre-amplified and measured using a digitizer \cite{kobayashi2020engineering}. The ESR spectrum was obtained by varying the external magnetic field, $B_0$, which confirmed the presence of P by the observation of the two hyperfine-split peaks due to the nuclear spin of $^{31}$P \cite{feher1959electron}. The $T_2$ was measured by setting the magnetic field to the value corresponding to the upper hyperfine-split P peak and using a standard Hahn echo pulse sequence \cite{kobayashi2020engineering}, where the pulse length (400 ns) and power were chosen such that the spins undergo a $\pi/2$ rotation for the first pulse. Due to the long $T_1$ of P donor electrons, the sample was illuminated with light of wavelength 1025 nm for 100 ms between each repetition of the pulse sequence in order to thermalize the donor spins faster \cite{tyryshkin2012electron}. 

To improve the enrichment process, the implantation of $^{28}$Si at various energies and fluences into $^\text{nat}$Si was simulated using TRIDYN, a binary collision Monte Carlo simulation package \cite{moller1988tridyn}. An initial interval spacing of $5$ {\AA} was chosen to be longer than the mean free path
but small enough to avoid artifacts from a coarse grid  
\cite{moller1985ion}. A high statistical quality was achieved using a precision of 0.02 to keep the maximum relative change of layer areal density per projectile to $<0.2$\%.

An additional $^\text{nat}$Si (UHPS Topsil) substrate was implanted with 45 keV $^{28}$Si$^-$ ions to a fluence of $2.63\times10^{18}$ cm$^{-2}$. A piranha (4:1 98\% H$_2$SO$_4$ : 30\% H$_2$O$_2$, 90 $^{\circ}$C) and RCA-2 (5:1:1 H$_2$O : 30\% H$_2$O$_2$ : 36\% HCl, 70 $^{\circ}$C) clean was then performed followed by a thermal anneal at 620 $^{\circ}$C for 10 min to facilitate SPE growth and a rapid thermal anneal at 1000 $^{\circ}$C for 5 s, suitable for donor activation, both performed in an Ar atmosphere.
 

The composition with depth of the 45 keV, $2.63\times10^{18}$ cm$^{-2}$ implanted sample after annealing was obtained with SIMS (IONTOF GmbH, TOF.SIMS 5). The Si isotopes were measured in positive polarity with a 1 keV O$_2^+$ beam used for sputtering and a 30 keV Bi$^+$ beam used for analysis. C and O impurities were measured in negative polarity with a 1 keV Cs$^+$ beam used for sputtering and a 30 keV Bi$^+$ beam used for analysis. 

TEM was used to determine the crystal quality of the enriched $^{28}$Si layer in the 45 keV, $2.63\times10^{18}$ cm$^{-2}$ implanted sample after annealing. Before TEM lamella preparation, the sample was coated with a thin carbon layer ($\sim$20 nm). To prepare the sample, a focused ion beam (FEI, Nova Nanolab 200) was used to grow a 300 nm thick layer of Pt via electron-beam assisted deposition. This was followed by a further 2.5 $\mu$m Pt layer deposited via a 30 keV Ga ion beam. A lamella was then extracted and thinned to a thickness of $\sim$100 nm, with a final polishing step performed with a 5 keV Ga ion beam. A TEM (FEI, Tecnai TF20) was used to take high-resolution cross-sectional images in which a 200 keV electron beam was transmitted down the [110] direction through the lamella to view the atomic arrangement.

\section{ESR at 3000 ppm $^{29}$Si}

 \begin{figure}[h]
\centering
\includegraphics[scale=0.48]{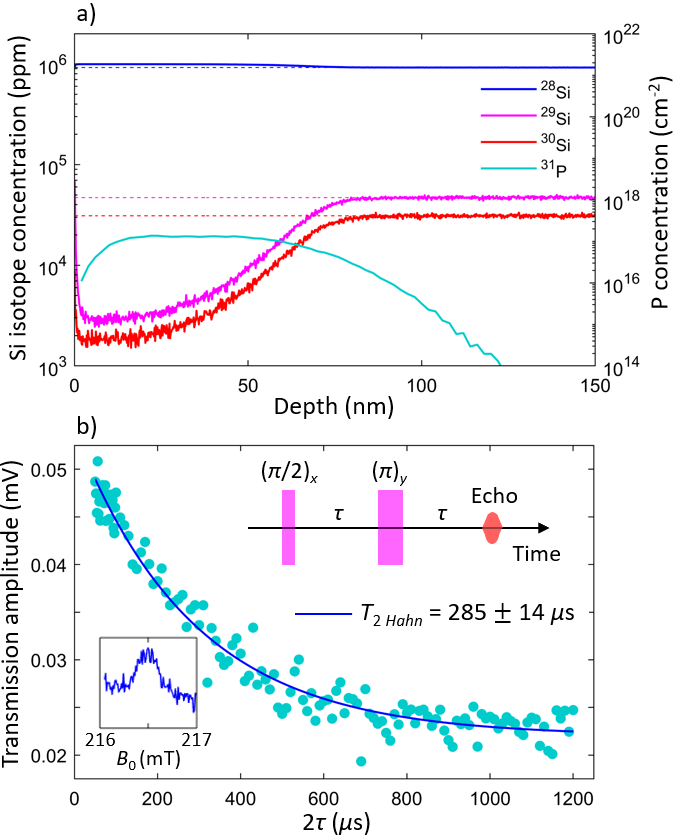}
\caption{Experimental data for the $^{28}$Si$^-$ (30 keV, $4\times10^{18}$ cm$^{-2}$) and $^{31}$P$^-$ (30 keV, $6.5\times10^{11}$ cm$^{-2}$ and 10 keV, $1.5\times10^{11}$ cm$^{-2}$) implanted sample after annealing. a) Left axis: SIMS depth profiles showing the concentration of the isotopes of $^\text{nat}$Si as a function of depth below the surface. Natural abundance is indicated with dashed lines. Right axis: SRIM simulation of the implanted P depth profile. b) Pulsed ESR measurement of the implanted P donors. The Hahn echo is fit with an exponential decay, indicative of a $^{28}$Si substrate, giving $T_2=285\pm14$ $\mu$s. The pulse sequence is shown in the top right and the upper hyperfine-split P ESR peak, collected with $\tau=5$ $\mu$s, is shown in the bottom left.}
\label{fig:T2}
\end{figure}

TRR was used to show the complete crystallisation of the $a$-Si layer in the 30 keV, $4\times10^{18}$ cm$^{-2}$ implanted sample via SPE; a thermally activated process with an intrinsic rate described by the Arrhenius relationship:
\begin{equation}
    v_i = v_0 \exp(E_A/k_B T)
\end{equation}
with $E_A=2.70$ eV and $v_0=4.64\times10^{16}$ \AA/s \cite{roth1990kinetics}. The SPE rate of the $a$-Si standard was calculated from the TRR curve to be 17.3 \AA/s, whereas that of the 30 keV, $4\times10^{18}$ cm$^{-2}$ implanted sample was 8.9 \AA/s. The increased level of impurities \cite{kennedy1977influence} and open-volume defects \cite{zhu1999instability} introduced by high fluence implantation slows the progression of the $a/c$ interface.

The isotopic enrichment level of the 30 keV, $4\times10^{18}$ cm$^{-2}$ implanted sample was measured using SIMS after annealing. The left axis of Fig. \ref{fig:T2}a shows the high fluence 30 keV $^{28}$Si implantation depleted $^{29}$Si and $^{30}$Si to concentrations of around 3000 ppm and 2000 ppm, respectively, in a surface layer $\sim$50 nm thick. The level of isotopic enrichment achieved with 30 keV is worse than that achieved with 45 keV, despite the higher implant fluence, as it is limited by self-sputtering of Si due to the sputter yield being greater than one. $^{31}$P$^-$ was implanted into the $^{28}$Si layer with a depth profile, simulated using SRIM, shown on the right axis of Fig. \ref{fig:T2}a. The maximum P concentration was confirmed to lie below the SIMS detection limit ($\sim2\times10^{17}$ cm$^{-2}$).

Pulsed ESR measurements were performed on the implanted P donors, as shown in Fig. \ref{fig:T2}b. The transverse relaxation time $T_2$ obtained using the Hahn echo pulse sequence gives $T_2=285$ $\mu$s $\pm14$ $\mu$s, which is comparable to previous reports \cite{chiba1972electron,tyryshkin2003electron} and longer than the coherence time obtained in $^\text{nat}$Si for similar doping concentrations ($\sim$100 $\mu$s \cite{chiba1972electron}). Moreover, the echo decay is mono-exponential which confirms that the instantaneous diffusion between the donor electrons is the dominant decoherence mechanism here, instead of the spectral diffusion caused by $^{29}$Si nuclear spins which dominates in $^\text{nat}$Si and adds an additional cubic term to the exponent \cite{chiba1972electron,tyryshkin2003electron,abe2010electron,tyryshkin2012electron}. These ESR measurements demonstrate that the P donors have been implanted into an isotopically enriched $^{28}$Si environment and successfully activated. In future, the concentration of P and residual $^{29}$Si can be reduced to improve coherence time.

\section{TRIDYN SIMULATIONS}

\begin{figure}[h]
\centering
\includegraphics[scale=0.32]{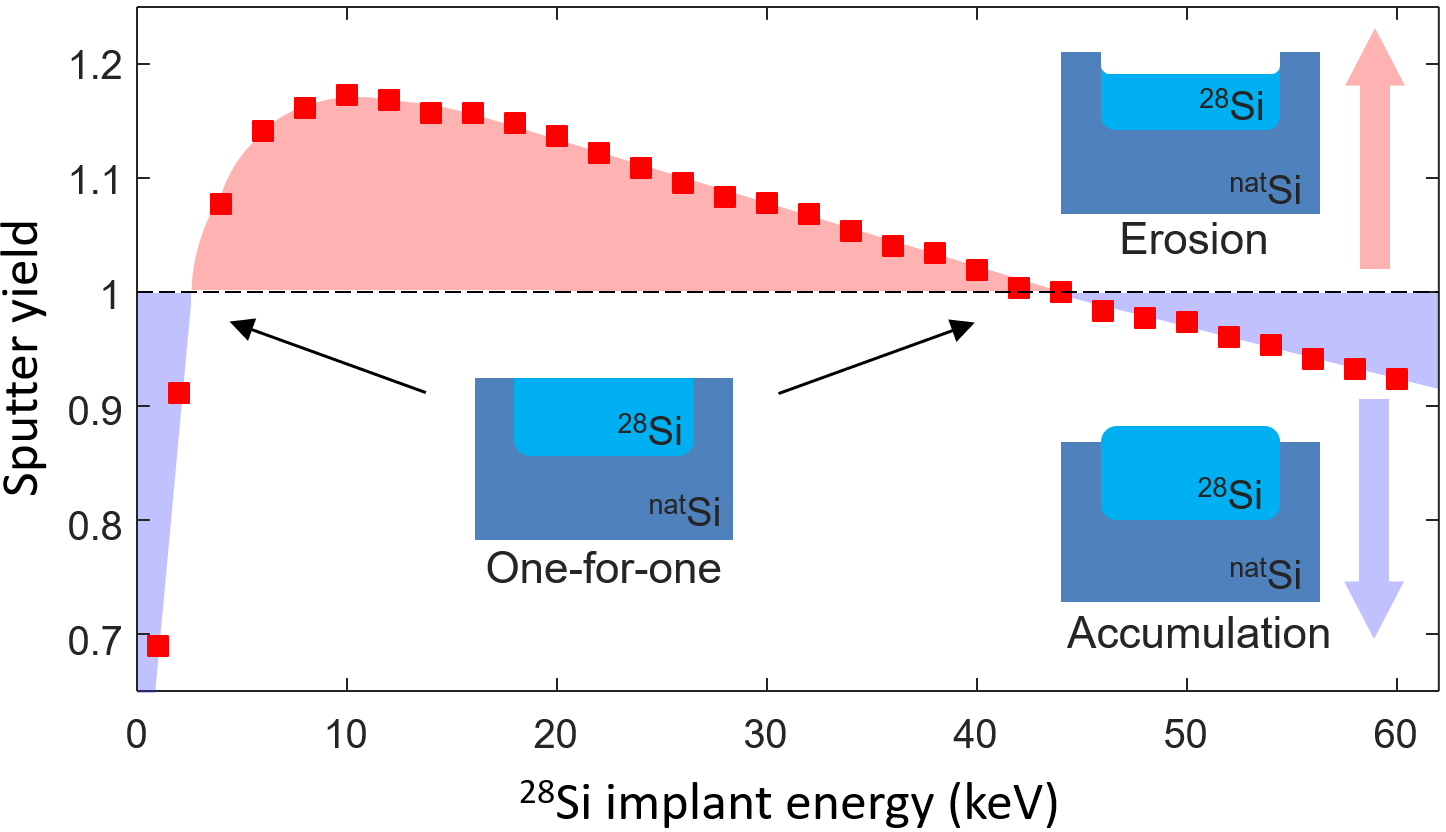}
\caption{TRIDYN simulation of the implantation of $^{28}$Si ions into $^\text{nat}$Si showing the sputter yield as a function of implant energy at a fluence of $1\times10^{17}$ cm$^{-2}$. Schematics of the post-implantation surface are shown in the erosion, one-for-one replacement and accumulation regimes.}
\label{fig:sputteryield}
\end{figure}

\begin{figure}[h]
\centering
\includegraphics[scale=0.6]{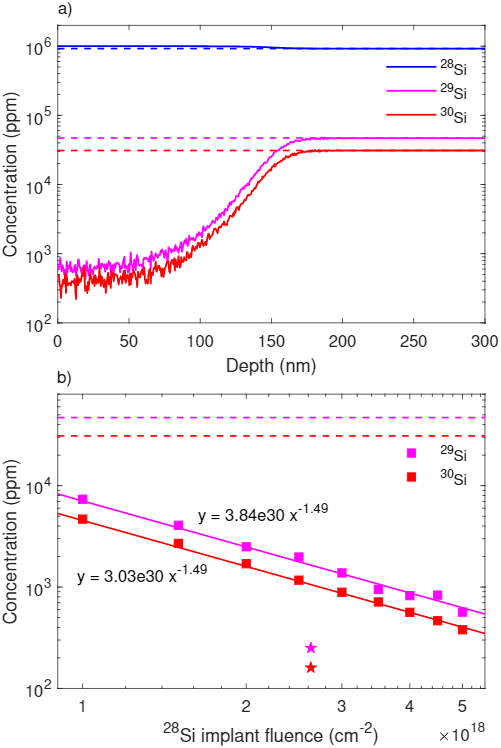}
\caption{TRIDYN simulation of the implantation of 45 keV $^{28}$Si ions into $^\text{nat}$Si. The dashed lines indicate natural abundance. a) The concentration of silicon isotopes as a function of depth after an implantation fluence of $5\times10^{18}$ cm$^{-2}$. b) The concentration of $^{29}$Si and $^{30}$Si at a depth of 20 nm below the surface as a function of implanted fluence. Lines of best fit are displayed for both isotopes. The star symbols represent the isotope concentrations achieved in this work and are extracted from Fig. \ref{fig:SIMS}.}
\label{fig:TRIDYNconc}
\end{figure}

To improve the enrichment process, TRIDYN simulations were performed. The sputter yield as a function of implantation energy was determined for an implantation of $1\times10^{17}$ cm$^{-2}$ $^{28}$Si ions at normal incidence, as shown in Fig. \ref{fig:sputteryield}. The sputter yield dependence on implantation energy shown here is in agreement with previous experimental Si sputter yields \cite{andersen1975heavy,frohlich1990angular} and theoretical fits \cite{eckstein2007sputtering}. An energy of $<3$ keV results in the deposition of $^{28}$Si onto the Si surface. If the sputter yield is greater than 1, the surface layer will be eroded faster than it can be isotopically enriched, resulting in a thin $^{28}$Si surface layer with reduced enrichment. $^{28}$Si ions with energies $>45$ keV are implanted deeper below the surface and sputtering is suppressed, resulting in accumulation. This is desirable for producing a thick layer of $^{28}$Si with a high level of enrichment, however, the surface will not be planar. A sputter yield of 1 is achieved at energies around 3 keV and 45 keV, both of which result in a planar surface; desirable for surface nanocircuitry fabrication. 45 keV was selected in order to produce a $^{28}$Si surface layer thicker than the qubit target depth of $\sim$20 nm in the one-for-one replacement regime and to optimize the transmission of the ion beam through the implanter. The sputter yield is independent of angle of incidence for angles below 10$^\circ$ for self-implanted Si \cite{frohlich1990angular} and so the TRIDYN simulations performed here at normal incidence are applicable for our experimental implants performed with a 7$^\circ$ tilt.

The depth profiles of Si isotopes in $^\text{nat}$Si after the simulated implantation of 45 keV $^{28}$Si at a fluence of $5\times10^{18}$ cm$^{-2}$ are shown in Fig. \ref{fig:TRIDYNconc}a. This shows that an isotopically enriched surface layer $\sim$100 nm thick is created. The resultant concentrations of $^{29}$Si and $^{30}$Si at a depth of 20 nm below the surface as a function of fluence of 45 keV $^{28}$Si are shown in Fig. \ref{fig:TRIDYNconc}b. This shows the trend of an increased isotopic purity resulting from an increased implant fluence. The isotope concentrations at a depth of 20 nm realised in this work with an implantation of $^{28}$Si$^-$ ions (45 keV, $2.63\times10^{18}$ cm$^{-2}$) as discussed below are indicated with star symbols in Fig. \ref{fig:TRIDYNconc}b.

\section{Depletion of $^{29}$Si to 250 ppm}

\begin{figure}[h]
\centering
\includegraphics[scale=0.6]{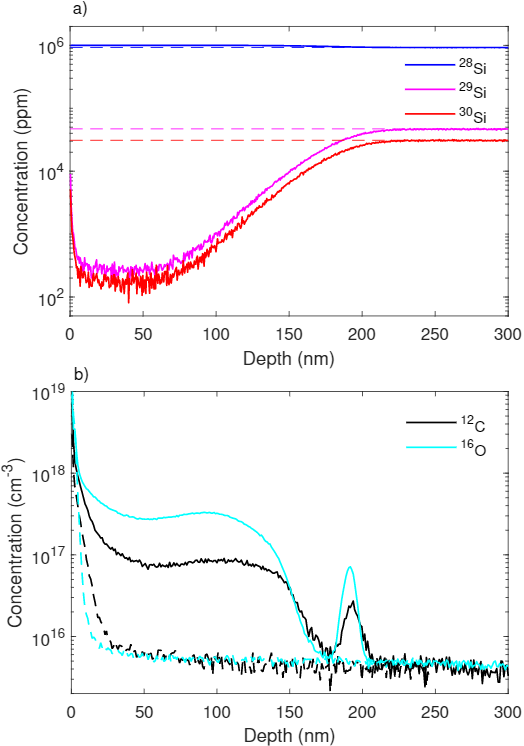}
\caption{Experimental SIMS depth profiles for the $^{28}$Si$^-$ (45 keV, $2.63\times10^{18}$ cm$^{-2}$) implanted sample. 
a) The concentration of the isotopes of $^\text{nat}$Si as a function of depth below the surface. Natural abundance is indicated with dashed lines. b) The concentration of the impurities $^{12}$C and $^{16}$O as a function of depth, calibrated with typical maximum background impurity levels expected in UHPS Topsil. Dashed lines indicate the SIMS depth profiles for the non-implanted substrate.}
\label{fig:SIMS}
\end{figure}

The concentrations of Si isotopes and $^{12}$C and $^{16}$O impurities in the $^{28}$Si$^-$ (45 keV, $2.63\times10^{18}$ cm$^{-2}$) implanted sample after annealing were measured with SIMS and are displayed as a function of depth below the surface in Fig. \ref{fig:SIMS}. Fig. \ref{fig:SIMS}a shows the high fluence $^{28}$Si$^-$ implantation depleted $^{29}$Si and $^{30}$Si to concentrations of around 250 ppm and 160 ppm, respectively, in a surface layer of thickness $\sim$100 nm. The shape of the isotope concentration profiles agree well with the TRIDYN simulation shown in Fig. \ref{fig:TRIDYNconc}a. A higher level of enrichment was achieved experimentally than predicted by TRIDYN, as shown by the star symbols in Fig. \ref{fig:TRIDYNconc}b. This could be due to the experimental sputter yield being slightly less than 1, leading to the accumulation of a thicker isotopically enriched layer, as evidenced by the depth where the isotope concentrations reach natural abundance: $\sim$180 nm for the TRIDYN simulation (Fig. \ref{fig:TRIDYNconc}a) and $\sim$220 nm for the experimental measurement (Fig. \ref{fig:SIMS}a). This accumulation was shown to result in lower $^{29}$Si and $^{30}$Si concentrations, demonstrated by TRIDYN simulations implanting $>$45 keV Si (not shown). The discrepancy in sputter yield, sensitive to the target surface binding energy, could be due to the impurity content of the substrate \cite{eckstein2007sputtering}. A smaller contribution could come from the uncertainty in the experimental implantation fluence. The residual $^{29}$Si concentration achieved here is around 3 times lower than that found in a commercially-produced $^{28}$Si wafer (Isonics) which, with 800 ppm $^{29}$Si \cite{itoh2014isotope}, has previously demonstrated increased coherence times of implanted donors \cite{muhonen2014storing}.
 
\begin{figure}[h]
\centering
\includegraphics[scale=0.42]{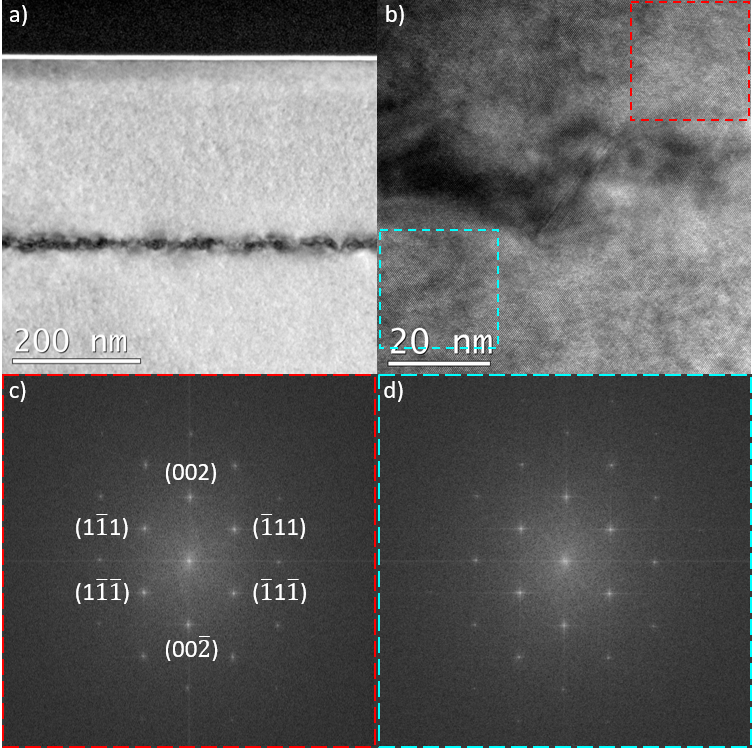}
\caption{ Cross-sectional TEM images of a lamella of the $^{28}$Si$^-$ (45 keV, $2.63\times10^{18}$ cm$^{-2}$) implanted sample after annealing. a) The end of range defects are visible as a dark band $\sim 290$ nm below the surface. b) High-resolution TEM image showing the successful repair of the crystal lattice in the implanted layer. Crystal diffraction patterns of c) the implanted region (highlighted in red) and d) the non-implanted region (highlighted in cyan).}
\label{fig:TEM}
\end{figure}

Fig. \ref{fig:SIMS}b shows the concentrations of $^{12}$C and $^{16}$O in an implanted region (solid lines) and in a non-implanted region of the substrate (dashed lines). The concentrations are increased above the background levels to around $1\times10^{17}$ cm$^{-3}$ for C and $3\times10^{17}$ cm$^{-3}$ for O by the process of high fluence implantation of $^{28}$Si$^-$ ions and subsequent annealing. The concentrations of these impurities were calibrated by assuming that the background levels at a depth of $\sim$300 nm, which match for the implanted and non-implanted regions, were $5\times10^{15}$ cm$^{-3}$; the maximum expected background contamination for UHPS Topsil quoted by the supplier. An increase in impurity levels, significantly above the background level in the surface $\sim$30 nm for C and surface $\sim$20 nm for O, is present in the non-implanted substrate. This accounts for some of the near-surface impurity content in the implanted region. If native SiO$_2$, typically $\sim$2 nm thick, was present, TRIDYN simulations (not shown) confirm the majority would be sputtered away during high fluence implantation, resulting in negligible O contamination from this source. The Si aperture reduced forward recoils of impurities, with no trace of heavy metals detected with high-resolution Rutherford backscattering spectrometry (not shown). The majority of the C and O contamination is proposed to be incorporated into the implanted layer from the imperfect vacuum, as seen before with high fluence implantation in a cryopumped target chamber \cite{singer1983absorption}. These levels of contamination are comparable to those present in Czochralski-grown Si ($4\times10^{17}-2\times10^{18}$ cm$^{-3}$ for O and $2\times10^{16}-4\times10^{17}$ cm$^{-3}$ for C \cite{liaw1981oxygen}) and indeed are shown to be low enough to allow for the successful crystallisation of the enriched layer by SPE. A peak in the concentration of C and O impurities occurs at around 190 nm below the surface of Si. This depth is significantly shallower than the depth of the end of range defects ($\sim$290 nm) visible with TEM as a dark band of dislocation loops, as shown in Fig. \ref{fig:TEM}a. The peak in impurity concentration at an intermediate depth between the surface and the end of range could be associated with the presence of open volume defects arising from vacancy clustering in this region and invisible to TEM. These open volume defects have been observed to act as gettering sites for impurities during annealing \cite{tamura1991mev,brown1998impurity}. Additionally, Fig. \ref{fig:SIMS}b shows preferential diffusion of C and O towards the surface, known to be a vacancy-rich region after ion implantation \cite{servidori1987retarded}.

The crystal quality of the $^{28}$Si$^-$ (45 keV, $2.63\times10^{18}$ cm$^{-2}$) implanted sample after annealing was determined using cross-sectional TEM, shown in Fig. \ref{fig:TEM}. End of range defects, visible as a dark band $\sim$290 nm below the surface in Fig. \ref{fig:TEM}a, indicate the location of the $a/c$ interface before annealing. The $a$-Si layer is extended to greater depths during continued ion bombardment above the Si amorphisation threshold \cite{claverie1988cross} (typically around $1\times10^{15}$ cm$^{-2}$ for keV Si ions \cite{mori2001optical}). The excess of interstitials at the end of range produced during ion implantation can evolve into dislocation loops during SPE regrowth of the $a$-Si layer \cite{jones1988systematic}. These dislocation loops are stable up to temperatures of 1100 $^\circ$C \cite{pan1997microstructural}, whereby they release self-interstitials into the surrounding substrate. This could cause undesired transient-enhanced diffusion \cite{stolk1997physical} of implanted P donor qubits in this enriched layer and so lower thermal budgets, supplied by low temperature SPE and rapid donor activation anneals, are preferred. Fast Fourier transforms (FFTs) of regions of the TEM image in Fig. \ref{fig:TEM}b were taken to give diffraction patterns indicating the crystal structure of the lamella. The diffraction pattern for the implanted region, shown in Fig. \ref{fig:TEM}c, indicates good crystal quality and matches that of the non-implanted $c$-Si substrate beneath, shown in Fig. \ref{fig:TEM}d. This shows the success of the crystallisation during post-implantation annealing. The contamination level introduced during the high fluence implantation is therefore low enough to avoid the formation of a polycrystalline $^{28}$Si layer, which would contain undesirable charge traps and dangling bonds at grain boundaries \cite{seager1985grain}. We expect that this single crystal layer of isotopically enriched $^{28}$Si will provide an ideal environment for implanted donor qubits, with high activation and long coherence times.

\section{Conclusion}

In conclusion, a Hahn echo measurement of P donors implanted into a $^{28}$Si layer with $\sim$3000 ppm $^{29}$Si, produced by high fluence implantation of 30 keV $^{28}$Si$^-$ ions, was fitted with a mono-exponential decay, suggesting an isotopically enriched $^{28}$Si donor environment. The extracted coherence time of $T_2=285\pm14$ $\mu$s is longer than that found with $^\text{nat}$Si for similar P concentrations. The residual level of $^{29}$Si was further decreased by implanting 45 keV $^{28}$Si$^-$ ions in the one-for-one sputtering regime. A high fluence ($2.63\times10^{18}$ cm$^{-2}$) implant of $^{28}$Si$^-$ ions at this energy into $^\text{nat}$Si results in a depletion of $^{29}$Si down to 250 ppm in a surface layer of thickness $\sim$100 nm, as measured with SIMS. The drastically reduced concentration of $^{29}$Si spin-1/2 nuclei in this isotopically enriched layer should further extend the coherence time of implanted donors beyond that achieved with commercial Isonics epilayers. Care was taken to limit the level of contamination introduced during high fluence implantation and concentrations were found to be below $1\times10^{17}$ cm$^{-3}$ for C and $3\times10^{17}$ cm$^{-3}$ for O, comparable to those in Czochralski-grown Si. The levels of contamination in this isotopically enriched $a$-Si layer are low enough to allow for successful crystallisation by SPE. The quality of the single crystal surface layer of $^{28}$Si was shown to be equivalent to the non-implanted region of the $c$-Si substrate using high-resolution TEM, in which the end of range defects were still visible after annealing. This work shows the high fluence implantation of $^{28}$Si$^-$ ions at energies around 45 keV is an effective method for isotopic enrichment which could be incorporated \textit{in-situ} into the fabrication of ion implanted donor spins in $^{28}$Si for quantum devices with increased coherence times.

\begin{acknowledgments}
This research was funded by the Australian Research Council Centre of Excellence for Quantum Computation and Communication Technology (CE170100012). We acknowledge the AFAiiR node of the NCRIS Heavy Ion Capability for access to ion-implantation facilities at EME, ANU, Silicon Quantum Computing for financial support and access to the electron spin resonance infrastructure, the Surface Analysis Laboratory, SSEAU, MWAC, UNSW for SIMS and the support of the International Atomic Energy Agency through the Cooperative Research Program number F11020 ``Ion beam induced spatio-temporal structural evolution of materials: Accelerators for a new technology era''.
\end{acknowledgments}

\nocite{*}
\bibliography{bibliography}

\end{document}